\documentclass[aps,twocolumn,prb,groupedaddress]{revtex4}
\usepackage{color,graphicx}
\usepackage{amsmath}
\usepackage{natbib}
\usepackage{longtable}

\begin{document}

\title{Magnetic transitions in CaMn$_7$O$_{12}$ : a Raman observation of spin-phonon couplings}


\author{C. Toulouse$^{1,2}$}
\altaffiliation[Present address :]{ Materials for Research and Technology - Ferroic Materials for Transducers, Luxembourg Institute of Science and Technology, 41 rue du Brill, L-4422 Belvaux, Luxembourg}
\altaffiliation[Corresponding Author :]{ constance.toulouse@list.lu}
\author{C. Martin$^3$}
\author{M-A. Measson$^{1,4}$}
\altaffiliation[Present address :]{Institut N\'eel, CNRS/UGA UPR2940, 25 rue des Martyrs BP 166, 38042 Grenoble cedex 9}
\author{Y. Gallais$^1$}
\author{A. Sacuto$^1$}
\author{M. Cazayous$^1$}
\affiliation{$^1$Laboratoire Mat\'eriaux et Ph\'enom\`enes Quantiques (UMR 7162 CNRS), 75205 Paris Cedex 13, France\\
$^2$Luxembourg Institute of Science and Technology, L-4422 Belvaux, Luxembourg\\
$^3$Laboratoire de Cristallographie et Science des Mat\'eriaux (CRISMAT (UMR 6508)), 14000 Caen, France\\
$^4$Institut N\'eel, CNRS/UGA UPR2940, 25 rue des Martyrs BP 166, 38042 Grenoble cedex 9}


\date{\today}

\begin{abstract}
The quadruple perovskite Calcium manganite (CaMn$_7$O$_{12}$) is a multiferroic material in which a giant magnetically-induced ferroelectric polarization has been reported, making it potentially very interesting for magnetoelectric applications. Here, we report the Raman spectroscopy study on this compound of both the phonon modes and the low energy excitations from 4~K to room temperature. A detailed study of the Raman active phonon excitations shows that three phonon modes evidence a spin-phonon coupling at T$_{N2}\simeq 50$~K. In particular, we show that the mode at 432~cm$^{-1}$ associated to Mn(B)O$_6$ (B position of the perovskite) rotations around the [111] cubic diagonal is impacted by the magnetic transition at 50~K and its coupling to the new modulation of the Mn spin in the (a,b) plane. 
At low energies, two large low energy excitations are observed at 25 and 47 cm$^{-1}$. The first one disappears at 50 K and the second one at 90~K. We have associated these excitations to electro-magneto-active modes. 
\end{abstract}

\maketitle

\section{Introduction}



\begin{figure}[b]
\centering\includegraphics[width = 0.48\textwidth]{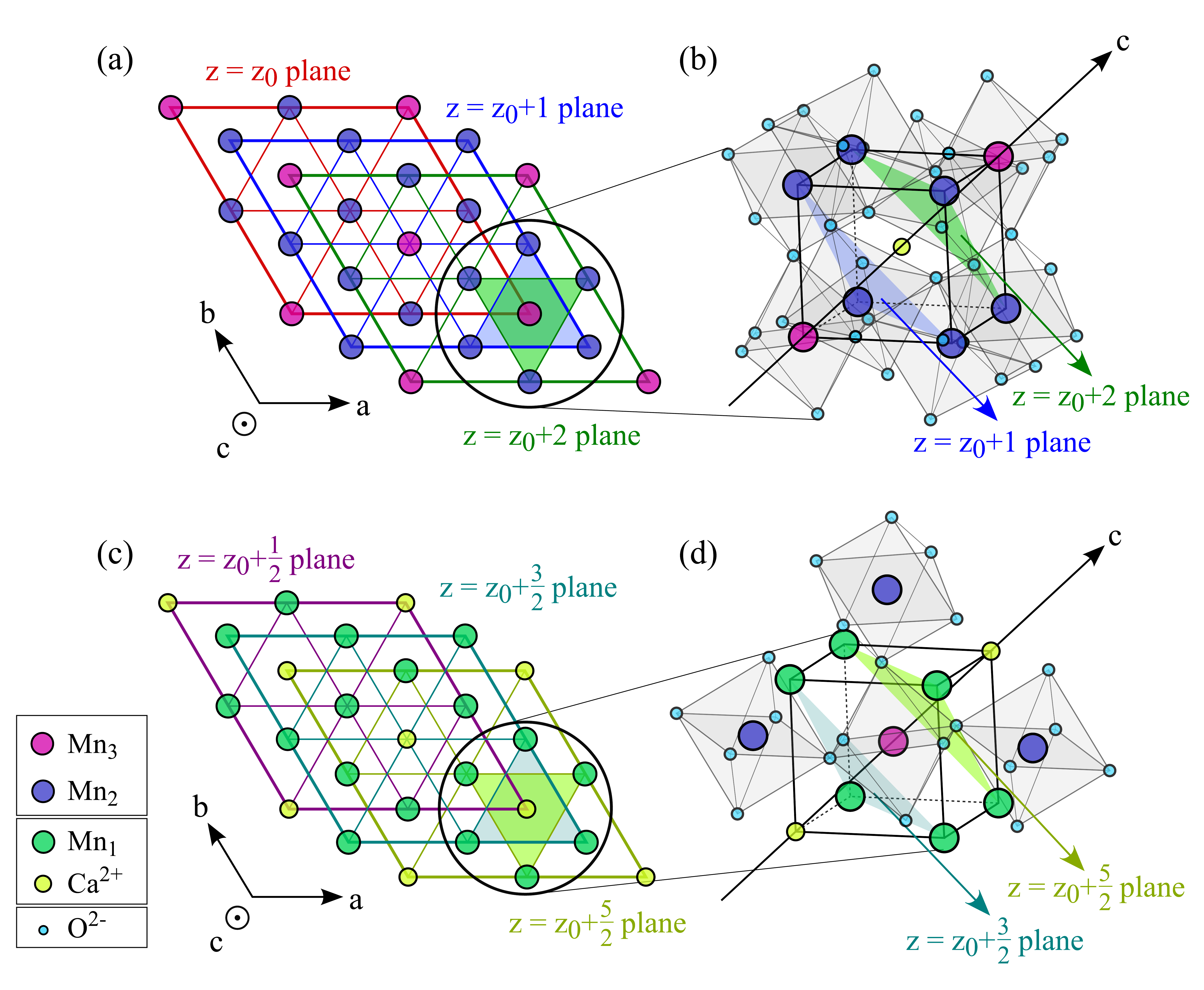}
\caption {Crystalline structure of CaMn$_7$O$_{12}$ shown in its trigonal ($R\bar{3}$ structure below 440 K. The different valence states (respectively Mn$^{3+}$ and Mn$^{4+}$) of the Mn$_2$ and Mn$_3$ sites are responsible for the appearance of an additionnal orbital ordering under 250 K.}
\label{structure}
\end{figure}

Since the discovery of multiferroic materials, compounds combining a huge ferroelectric polarization and an important magnetoelectric coupling have been the goal for research in material synthesis.\cite{Eerenstein, Zhao, Senff}
Magnetoelectric multiferroic materials, combining ferroelectric and magnetic order in the same phase, can be classified in two categories\cite{Khomskii} : type I materials in which the two orders appear independently at different transition temperature, exhibiting in general a low coupling between the two orders\cite{Wang}, and type II materials where the ferroelectricity is induced by an another order (magnetic, orbital, electronic)\cite{Cheong, Sergienko}. The latter generally exhibit a much more important magnetoelectric coupling but have a small ferroelectric polarization ($\sim$ 100 $\mu.C.m^{-2}$) appearing at very low temperature (below 30 K).

\begin{figure}[tb]
\centering\includegraphics[width = 0.4\textwidth]{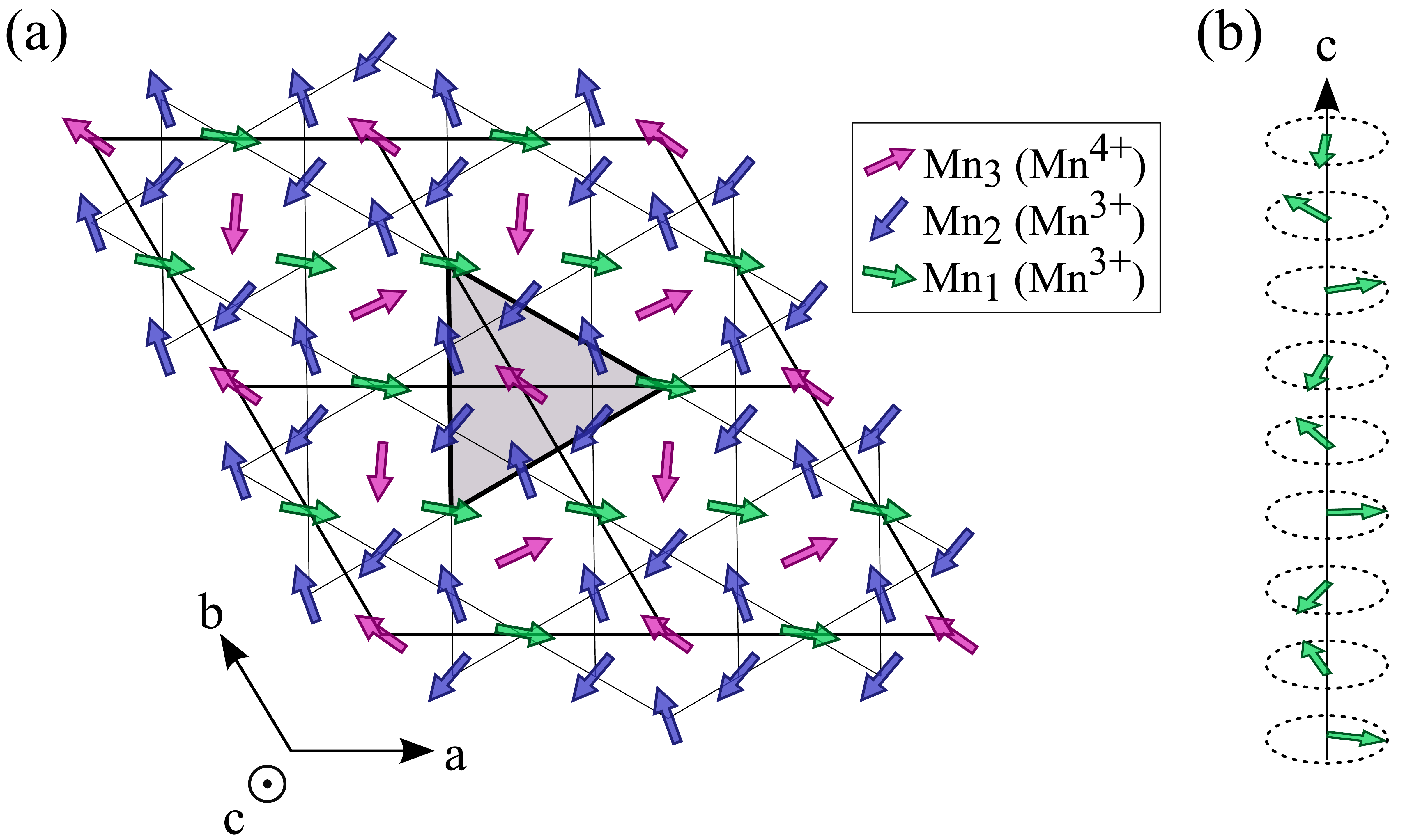}
\caption {Antiferromagnetic structure of CaMn$_7$O$_{12}$ appearing below 90 K (AFM$_I$ phase). The spin of the Mn ions order antiferromagnetically in the ($a$,$b$) planes with superimposed long range spin spirals along the $c$ axis.}
\label{mag_structure}
\end{figure}

The exception is the quadruple calcium manganite, CaMn$_7$O$_{12}$, which has been synthesized as monocrystal of good quality only recently, which exhibits giant spin-induced polarization of 2870~$\mu.C.m^{-2}$ in cycloidal AFM phase below 90 K.

Although this observation is supported by both experimental \cite{Zhang2011, Johnson2012} and theoretical \cite{Perks2012, Lu2012} works, the giant ferroelectricity in CaMn$_7$O$_{12}$ is currently under discussion. The high values of the polarizations reported in literature have been recently associated to extrinsic thermal effect rather to an intrinsic ferroelectricity\cite{Terada2016} although a very small ferroelectric polarization of 0.2~$\mu.C.m^{-2}$ is still reported to subsist under 46~K.

Under 440 K, CaMn$_7$O$_{12}$ crystallizes in the trigonal $R\bar{3}$ space group \cite{Przenioslo2002}. Its structure is a quadruple ABO$_3$ rhombohedrally distorted perovskite structure with Ca and Mn ions (Mn$_1$ sites) sitting in the A sites and Mn ions (Mn$_2$ and Mn$_3$ sites sitting in the B sites of the structure (see Fig.~\ref{structure}). This structural transition is accompanied by metal-insulator transition and charge ordering.   The B-site Mn ions order into Mn$_3^+$ and Mn$_4^+$ with a 3 : 1 ratio.\cite{Lim2018}
The subsequent structural helicoidal modulation of the Mn-Mn bond angles is incommensurate and propagates along $q_C \simeq (0,0,2.077)$.\cite{Perks2012}

The manganese being a magnetic specie, the Mn$^{3+}$ ($3d^4$, $S = 2$, $L = 2$) and Mn$^{4+}$ ($3d^3$, $S = \frac{3}{2}$, $L = 3$) are responsible for the appearance of two successive magnetic orderings \cite{Sanchez-Andujar2009}. At T$_{N1}$=90 K, the Mn spins orient in the ($a$,$b$) planes as shown in Fig.~\ref{mag_structure} and are coupled antiferromagnetically between neighboring planes (with 124$^{\circ}$ angles), forming incommensurate magnetic spirals along the 3-fold $c$ axis.
These spirals are coupled to the orbital ordering : they lie along the same direction with twice the wavelength of the orbital spirals. The propagation wavevector of the spin spiral is equal to $q_M=(0, 0, 1.037)$ \cite{Perks2012} in hexagonal notations or $k_1=(0, 1, 0.963)$ in pseudo-cubic system \cite{Johnson2012}. For the rest of the article, we will denote this phase AFM$_\mathrm{I}$.
Below T$_{N2}\simeq 50$ K, a second magnetic transition towards the AFM$_\mathrm{II}$ phase occurs and a new modulation of the Mn spins in the ($a$,$b$) planes appears corresponding to a `beating' of the previous AFM$_\mathrm{I}$ order. The propagation wavevector of the spin spiral gives rise to two near propagation wave-vectors $k_{\pm}=k_1\pm(0, 0, \delta)$.

The possible coupling between these magnetic orderings and a strong electric polarization arises questions regarding the mechanisms at stake in this compound. Moreover, it has been shown that the lattice degrees of freedom are actively involved in the orbital-ordering and magnetic mechanism.\cite{Du2014, Souliou2014} The study of the phonon modes and of the low energy excitations at the magnetic transitions by Raman spectroscopy is an efficient way to investigate the possible couplings between the different multiferroic orders. Up to now, very few Raman experiments have been reported\cite{Du2014, Iliev2014, Nonato2014} on CaMn$_7$O$_{12}$. We present here a Raman investigation on both the low energy magnetic excitations and the phonon modes, with a focus on the temperature dependencies around the two magnetic transitions. At low temperature, it has been shown in powdered samples that the magnetostriction plays a role in the second magnetic transition at 50~K, and hence could be linked to our observations\cite{Nonato2014}. We have observed two additional excitations at low energies and could link them to the two magnetic transitions with their temperature behaviour. These two excitations have been identified as the Raman signature of CaMn$_7$O$_{12}$'s electromagnons.

\section{Experimental Details}

\begin{figure}[bt]
\centering\includegraphics[width = 0.5\textwidth]{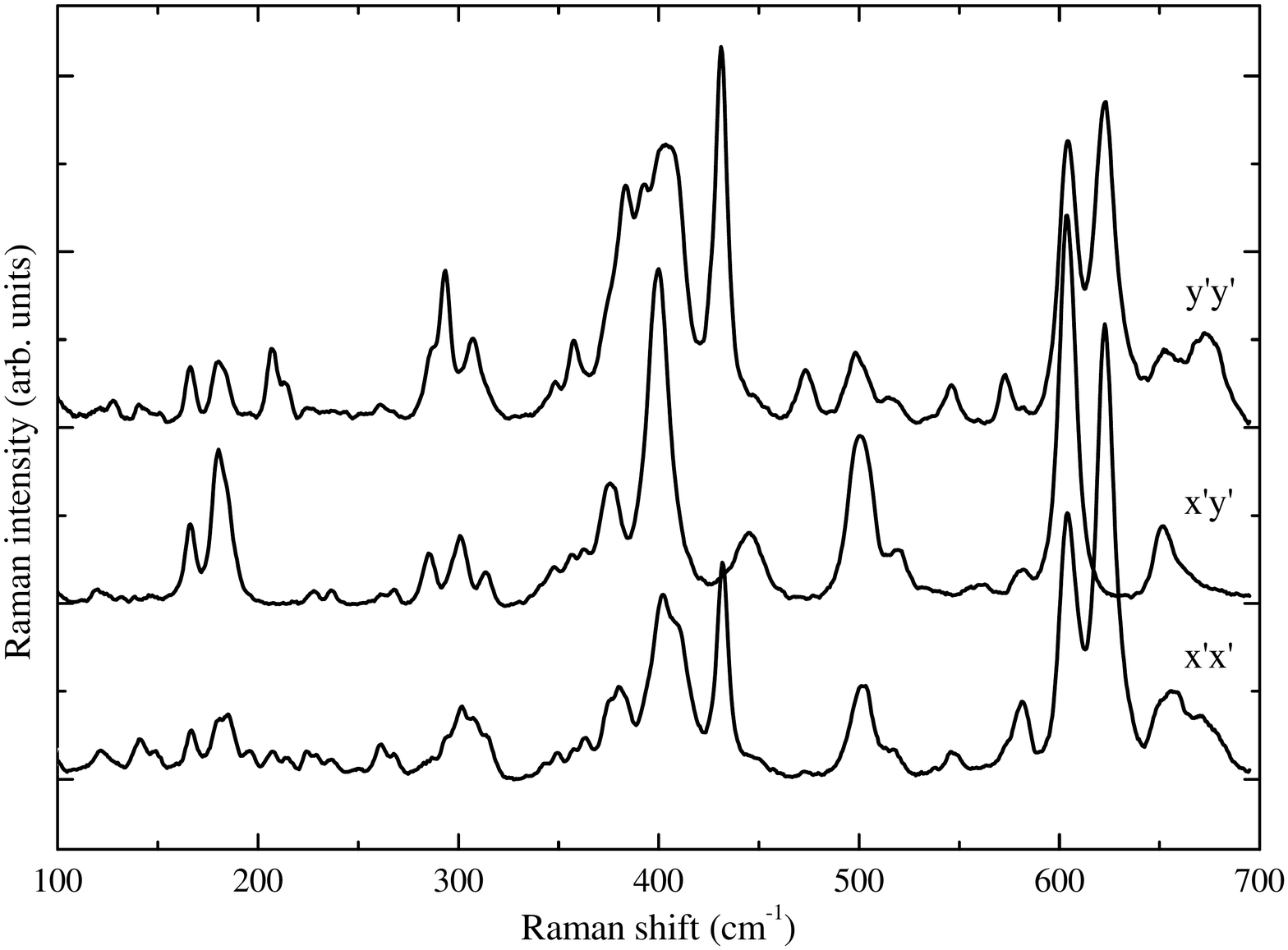}
\caption {Low temperature (10K) Raman spectra of phonon modes using  cross-polarization (x'y') and with parallel-polarization (x'x' and y'y') configurations. The associated assignements are derived from \onlinecite{Iliev2014}.}
\label{phonons_selec}
\end{figure}



The CaMn$_7$O$_{12}$ samples studied here are high quality single crystals of approximately 60$\times$60$\times$60 $\mu m$ size. They have been grown by flux starting from a (CaCl$_2$:MnO$_2$) mixture with weight ratio (1:3) heated at 850$^{\circ}$C for 24 hours and then cooled down at a rate of 5$^{\circ}$C per hour. Bulk properties measurements and X-ray diffraction have been performed\cite{Johnson2012, Johnson2016} and showed no twinning at room temperature. 


The Raman spectroscopy measurements were recorded using a triple substractive T-64000 Jobin-Yvon spectrometer with a cooled CCD detector. The spectra were acquired using a 561.3 nm laser line from an Oxxius-Slim solid state laser, filtered both spatially and in frequency. Measurements
between 10 and 300 K with an error bar of $\pm 2 K$ have been performed using an ARS closed-cycle He cryostat. The instrumental response is a Gaussian with a FWHM of about 0.5 cm$^{-1}$. The conditions of sample preparation are important and only raw samples present a small background allowing to observe the low energy excitations.

\section{Results and discussion}

\subsection{Lattice modes}

\begin{table}[!hb]
\centering
\caption{Phonon modes measured by Raman spectroscopy compared with the modes measured in ref. \onlinecite{Iliev2014}. Note that the structure modulation below 250 K and in addition the breaking of space symmetry in the magnetic phase (IR modes become also Raman active) allow a higher number of modes.}
\renewcommand{\arraystretch}{1.2}
\begin{tabular}{c|cc}
\hline
\hline
Our work & \multicolumn{2}{c}{Iliev \& al.}\\
\hline
Energy (cm$^{-1}$) & Energy (cm$^{-1}$) & Mode assignment\\
\hline
78 & -- & -- \\
99 & -- & -- \\
122 & -- & -- \\
140 & -- & -- \\
150 & -- & -- \\
168 & 167 & Eg \\
180 & 179/180 & Eg \\
186 & 185 & Ag \\
207 & 207 & Ag \\
215 & 211/215 & Ag \\
227 & -- & -- \\
238 & -- & -- \\
262 & -- & -- \\
269 & -- & -- \\
287 & 285/286 & Eg\\
294 & -- & -- \\
301 & 301 & Eg \\
309 & 306 & Eg \\
316 & 314 & Eg \\
352 & -- & -- \\
359 & -- & -- \\
364 & -- & -- \\
375 & 376 & Eg \\
380 & -- & -- \\
385 & 390 & Eg \\
403 & 400 & Eg \\
410 & 412 & Ag \\
432 & 428/432 & Ag \\
452 & -- & -- \\
472 & 473/474 & Ag \\
493 & 497 & Eg \\
501 & 501 & Eg \\
518 & -- & -- \\
533 & -- & -- \\
546 & -- & -- \\
560 & -- & -- \\
580 & 597 & Eg \\
605 & 604 & Eg \\
622 & 623 & Ag \\
652 & 651 & Eg \\
\hline
\hline
\end{tabular}
\label{modes_Raman}
\end{table}

\begin{figure}[tb]
\centering\includegraphics[width = 0.5\textwidth]{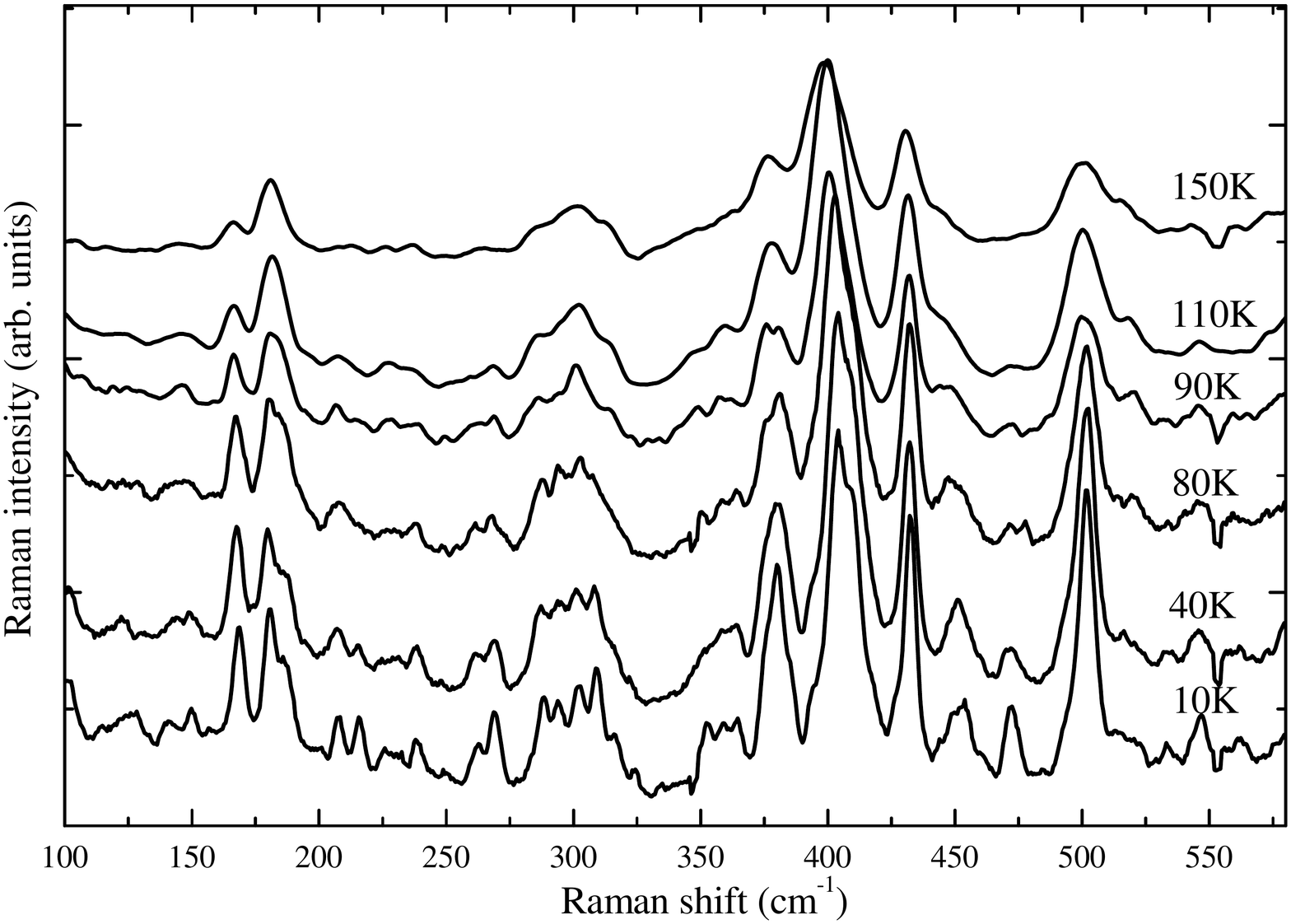}
\caption {Temperature dependence of the Raman spectra from 10~K to 150~K in parallel polarization configuration. No difference in number of phonon modes is observed.}
\label{phonons_T}
\end{figure}

CaMn$_7$O$_{12}$ crystallizes in the $R\bar{3}$ trigonal space group (148) and the total number of phonon modes at the $\Gamma$ point is $\Gamma_{\mathrm{TOT}}=6 A_g + 6 E_g + 14 A_u + 14 E_u$. Twelve Raman active phonon modes are expected for first order processes at the Brillouin zone center : $\Gamma_{\mathrm{Raman}}=6 A_g + 6 E_g$.\cite{Iliev2014} Using different scattering configurations, it's possible to choose the mode to activate. Here we used incident wave vector anti-parallel to the scattered one (backscattering configuration). The E$_g$ and E$_g$+A$_g$ modes are activated with 
cross-polarizations (x'y') and with parallel-polarizations (x'x' and y'y'), respectively. The x' and y' notations correspond to the directions parallel to the [110] and [$\bar{1}$10] quasicubic directions, respectively. 

\begin{figure}[t]
\centering\includegraphics[width = 0.4\textwidth]{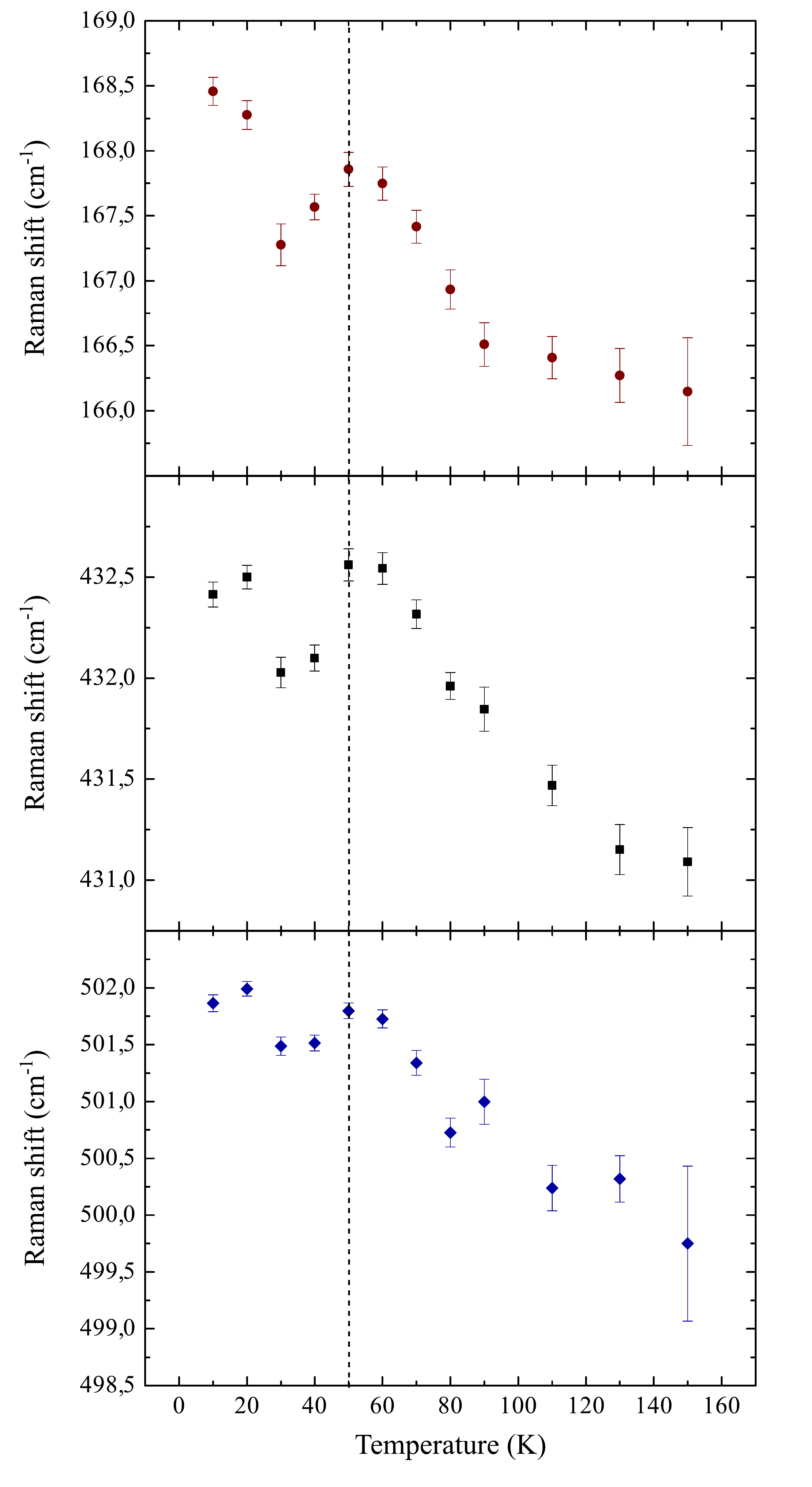}
\caption {Temperature dependence of 168.5, 432.5 and 502 cm$^{-1}$ phonon modes's wavenumber. The dotted line correspond to the T$_{N2}$ magnetic transition temperature.}
\label{phonon_behav}
\end{figure}

Figure~\ref{phonons_selec} shows low temperature Raman spectra obtained at 10~K. We have reported the frequencies of the A$_g$ and E$_g$ modes in Table \ref{modes_Raman}. Our results are compared to the measurements of Ref. \onlinecite{Iliev2014}. We have been able to identify 37 modes compared to Ref. \onlinecite{Iliev2014}. Notice that it was not possible with the Raman selection rules to determine the symmetry of low intensity peaks. It is clearly visible that more than 12 Raman modes are observed. Although this large number of modes could be explained by a low quality of the single-crystal samples, we discarded that possibility due to several characterizations (X-ray and neutron diffraction) that have demonstrated their quality. Moreover, there is no evidence of nano-domains in the characterization measurements\cite{Johnson2012, Perks2012, Johnson2016}. Additionally, the modes observed in the Raman spectra seem too sharp and their energies are too low to be due to second order processes. One can also notice that the Raman spectra shown in Iliev et al.\cite{Iliev2014} exhibit an equivalent number of modes for crystals from a different source. 

To study the coupling between the lattice and the magnetic orders, we measured the temperature dependences of the phonon modes.  
Figure~\ref{phonons_T} presents the Raman spectra from 4~K to 150~K above 100~cm$^{-1}$. 
We can observe that the number of phonon modes seems to increase around the magnetic transitions at 90~K and 50~K. It has been shown by Slawinski et al.\cite{Slawinski2012} that the magnetic transitions are associated to incommensurate modulation of the atomic positions. All phonon modes might become both Raman and infrared active due to the modulation and in particular out-of-center phonon modes might become Raman active and visible in our spectra. Part of the additional modes might be related to the symmetry rule breaking due to the incommensurate transition but it might be also necessary to consider a new space group for the structure at low temperatures. 

\begin{figure}[tb]
\centering\includegraphics[width = 0.5\textwidth]{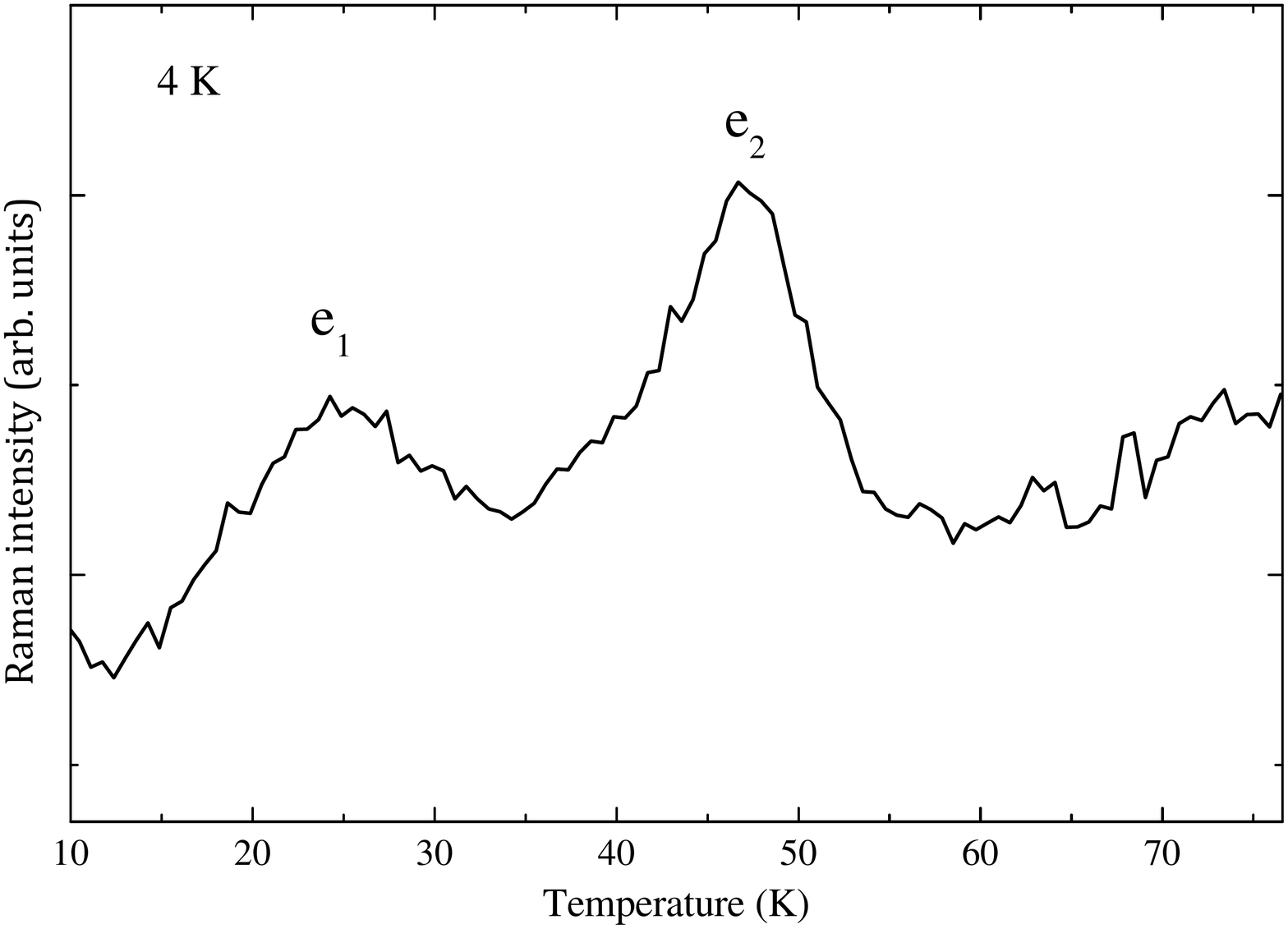}
\caption {Low energy Raman excitations measured at 4 K in CaMn$_7$O$_{12}$ in parallel polarization configuration.}
\label{RefLT}
\end{figure}

Figure \ref{phonon_behav} shows the behavior in temperature of three phonon modes : the Ag mode at 432~cm$^{-1}$ and the two Eg modes at 168 and 502~cm$^{-1}$. All the phonon frequencies soften due to the dilation of the unit cell when temperature increases.
These phonon modes present an abrupt change in their frequencies around the lowest magnetic transition temperature T$_{N2}\simeq 50$~K that can be interpreted as the fingerprint of a spin-phonon coupling at this magnetically-ordered phase.
We can notice that the first magnetic transition at 90 K can induce changes in the Raman mode frequencies due to the direct coupling between the magnetic ordering and lattice since there is no anomaly in the lattice parameters at this magnetic transition. No significant changes in the mode frequencies has been measured at this transition. 
However, at the second magnetic transition the lattice parameters change significantly and the spin-phonon coupling anomalies we are measuring at 50~K could be due to the magnetostriction effect evidenced by Nonato \textit{et al.}\cite{Nonato2014}.
The Ag mode at 432~cm$^{-1}$ has been associated to Mn(B)O$_6$ (B position of the perovskite) rotations around the [111] cubic diagonal.\cite{Iliev2014}
This mode is quite important because it is the only Ag mode to survive at high temperatures in the Im$\bar{3}$ phase  with disordered Jahn-Teller distortions.\cite{Iliev2014} The abrupt change of slope at 50~K shows that the new modulation of the Mn spin in the (a,b) plane, possibly driven by the magnetostriction effect, controls this mode. 
Notice that Nonato \textit{et al.}\cite{Nonato2014} have observed anomalies in the phonon frequencies in polycrystalline compounds. They observed an inversion of the frequency slope around 50 K for a peak at 378.8~cm$^{-1}$ (376~cm$^{-1}$ in our measurements). We have also detected a shift but within our experimental error bars. We have not detected the strong shift detected for the 625~cm$^{-1}$ peak around 60 K and 90 K. Such a discrepancy might originate from the samples (monocrystal versus polycrystal).

\subsection{Low energy excitations}

Figure \ref{RefLT} shows the low energy Raman spectrum obtained at 4~K in parallel polarization configuration.
Two large low-energy Raman excitations labeled e$_1$  and e$_2$ are visible around 25 and 47 cm$^{-1}$, respectively.

\begin{figure}[b]
\centering\includegraphics[width = 0.5\textwidth]{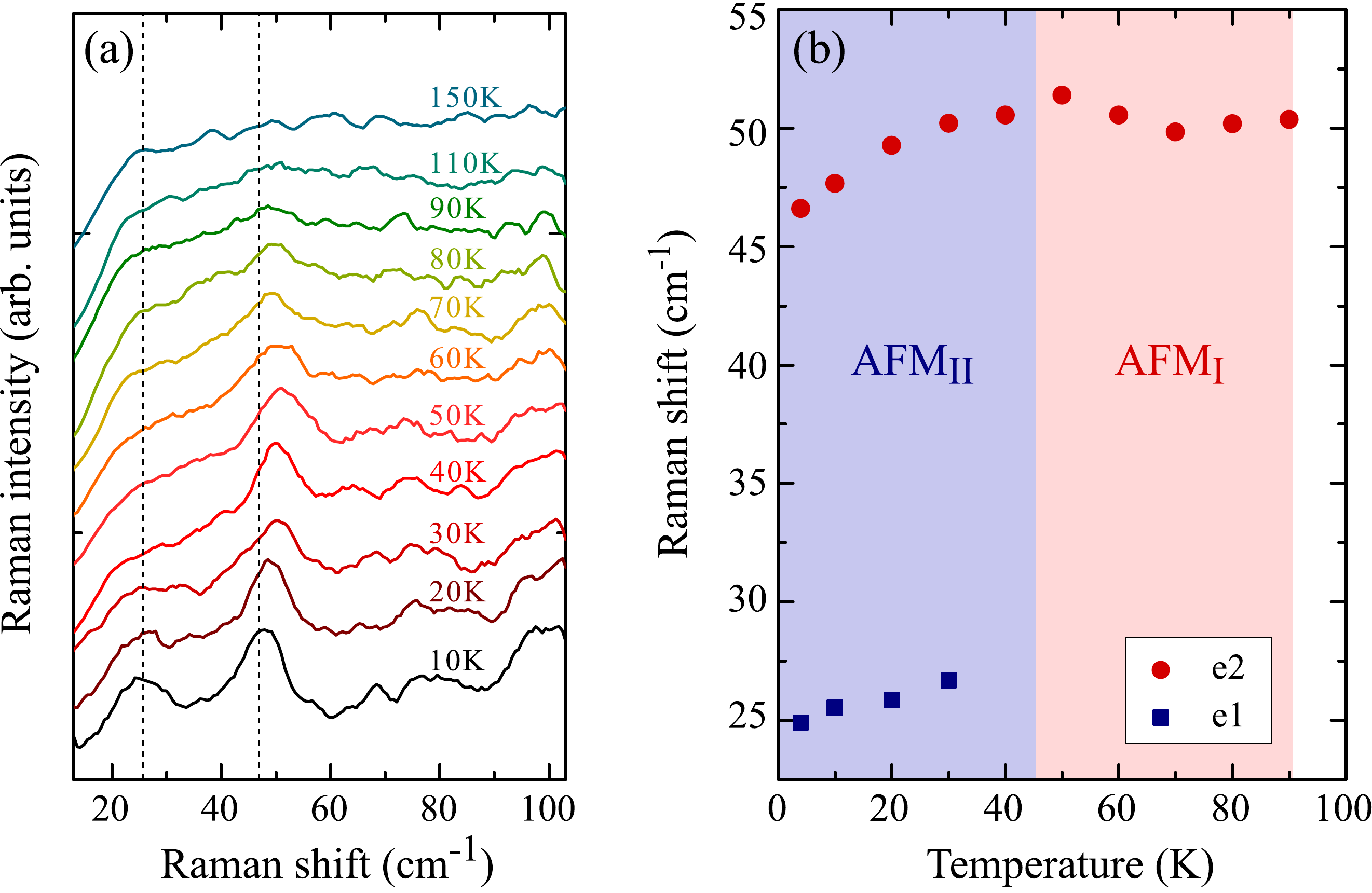}
\caption {Temperature dependence of : (a) low energy Raman spectra of CaMn$_7$O$_{12}$ between 10 and 150 K, (b) the Raman shift of the e$_1$ (square) and e$_2$ (circle) low energy excitations as a function of the temperature.} 
\label{DepT}
\end{figure}

To have a better insight on the nature of these two excitations, the temperature dependence of the low energy Raman spectra have been recorded between 4~K and 150~K in Fig.~\ref{DepT}(a).

In Fig.~\ref{DepT}(b), the e$_1$ lower energy excitation disappears at the magnetic transition (T$_{N1}\simeq$50 K) whereas the e$_2$ excitation at 47~cm$^{-1}$ survives above this temperature to disappear at the magnetic transition T$_{N2}\simeq$90 K.
The e$_1$ excitation is thus connected to the modulation of the magnetic order occurring in the AFM$_\mathrm{II}$ state and e$_2$ excitation is related to  the AFM$_\mathrm{I}$ ordering of the Mn spins in the $(a,b)$ planes with spiral propagation along the 3-fold axis. Remember that this transition is reportedly responsible for the appearance of the ferroelectric order.
Notice that the frequency of the e$_2$ excitation downshifts below the second magnetic phase. This behavior indicates a sensitivity of this excitation to the magnetic modulation occurring at 50 K.
Our observations are in agreement with the measures published by Kadlec et al.\cite{Kadlec2014} where they measured using Infrared spectroscopy electromagnons excitations at the same energies. 
On the contrary, these modes don't remain active in the paramagnetic phase and we have not observed paraelectromagnons. 
A soft mode has also been reported at around 80cm-1 at low temperatures (10 K)\cite{Du2014, Souliou2014}. Although we didn’t report this mode in our study, the signal over noise ratio of our low energy Raman data is not allowing us to exclude with certainty the observation of this mode.

\section{conclusion}

In summary, the high number of phonon modes observed in CaMn$_7$O$_{12}$ rises the question of the space group of this compound at low temperature. 
The temperature dependance of the lattice excitations in CaMn$_7$O$_{12}$ show the role plays by magnetostriction coupling in this compound.
In particular, the Ag mode at 432~cm$^{-1}$  associated to Mn(B)O$_6$ rotations around the [111] cubic diagonal is modified the modulation of the Mn spin in the (a,b) at 50 K. We were able to measure low energy Raman excitations in CaMn$_7$O$_{12}$ and observed two large excitations around 25 and 47 cm$^{-1}$. We show that they are linked to the two different magnetic orderings occurring below 90 and 50 K and are interpreted as the Raman signature of CaMn$_7$O$_{12}$'s electromagnons.

\section*{Acknowledgments}
This work was supported in part by the French National Research Agency (ANR) through DYMMOS project and by the General Directorate for Armament (DGA).


\end{document}